\documentstyle[11pt]{article}

\newcommand{\beq}{\begin{equation}}
\newcommand{\eeq}{\end{equation}}
\newcommand{\p}{\partial}
\newcommand{\bfe}{\mbox{\bf e}}
\newcommand{\bfi}{\mbox{\bf i}}
\newcommand{\bfj}{\mbox{\bf j}}
\newcommand{\bfk}{\mbox{\bf k}}
\newcommand{\bfr}{\mbox{\bf r}}
\newcommand{\bfP}{\mbox{\bf P}}
\newcommand{\bfsigma}{\mbox{\boldmath $\sigma$}}
\newcommand{\stilde}{\widetilde{s}}
\newcommand{\etilde}{\widetilde{e}}
\newcommand{\atilde}{\widetilde{a}}
\newcommand{\btilde}{\widetilde{b}}
\newcommand{\sigtilde}{\widetilde{\sigma}}
\newcommand{\disp}{\displaystyle}

\begin{document}

\begin{center}

{\Large{\bf Electromagnetism and Gravitation}}\\

\vspace{.5in}

{\large {\bf Kenneth Dalton}} \\
Post Office Box 587 \\
Mammoth Hot Springs \\
Yellowstone Park WY 82190, U.S.A. \\

\vspace{1.in}

{\bf Abstract }

\end{center}
\vspace{.25in}

Modern experiment has shown that bulk matter exhibits quantum structure and
that this structure is electromagnetic in character.  Therefore, we treat the
electromagnetic field as the source of gravitation.  In this way, the theory of
gravitation becomes consistent with the quantum theory of matter, which holds
that electric charge (or `generalized charge') is the most fundamental
attribute of matter.

The following predictions of the theory may lie within reach of present-day
methods in astronomy:
\begin{itemize}

\item[(1)] any massive body generates a time-dependent gravitational field;

\item[(2)] there is a linear correlation between the gravitational red-shift of
a stellar source and the energy of cosmic rays emitted by that source, given by
$ {\Delta \nu}/{\nu_0} = {\mbox{\rm energy\, (eV)}}/10^{27} $;

\item[(3)] the maximum energy of cosmic rays is $ 10^{27} $ eV;

\item[(4)] this limit is associated with an infinitely red-shifted stellar
object---an ``electrostatic black-hole''---at the potential
$ c^2/G^{1/2} = 10^{27} $ volts.

\end{itemize}
Finally, the theory predicts that the gravitational potential near any charged
elementary particle is many orders of magnitude greater than the Newtonian
value.

\clearpage

\section*{\large {I. Introduction}}

\indent

In the theory which follows, electromagnetism is taken to be the source of
gravitation.  The foundation for this hypothesis lies in the observed nature of
matter itself.  Modern experiment has shown that matter can no longer be
described, in any fundamental way, by an amorphous `density of mass.'  Rather,
the experiments reveal an intricate quantum structure which is largely
electromagnetic in character.  This profound change in our knowledge of matter 
has met with no corresponding change in the theory of the gravitational field.
Consequently, a perplexing estrangement has replaced the close historical tie
between gravity and matter.  \footnote{\small Evidence for this may be found in
any book on modern physics.  On the one hand, there are the strong, 
electromagnetic, and weak interactions between the elementary particles of 
matter; on the other hand, some 40 orders of magnitude removed, there is the 
Newtonian gravitational interaction of `bulk' matter. }
It is toward restoring this close tie that we propose to trace the origin of
gravitation to the quantum of electric charge.

A development of this nature cannot take place within the context of Einstein's
theory of general relativity.  According to that theory, electrostatic energy
produces a repulsive gravitational field.  (This follows from the Reissner-
Nordstrom solution of Einstein's field equations, for a free charged particle,
at rest.)  In order to arrive at the attractive gravitational field of
experience, we must reformulate the problem of space and time.  To this end,
we abandon the four-dimensional vector basis of general relativity.  At each
point in space-time, we introduce a scalar basis $e_0$ and a three-dimensional
vector basis $ {\bfe}_i,\; i = 1,2,3 $.  We begin by expressing the Lorentz
transformation in terms of this scalar, three-vector basis.  This was first
accomplished by L.~Silberstein in 1912 [1].  We then go on to eliminate
relative motion from the theory, by suitably restricting the transformations
of space and time.  This gives rise to a covariant expression which is
identified with the gravitational field.  Thus, according to our theory, a
gravitational field induces non-uniformity in space and time which `breaks the
symmetry' of Lorentz invariance.  This field has well-defined energy, momentum,
and stress at each point in space.  Moreover, it has tensor character and
cannot be arbitrarily transformed to zero.  Finally, it interacts with
electromagnetism via a tensor force and thereby justifies our unified treatment
of the field.

\clearpage

\section*{\large {II.  Transformations of Space and Time}}

\indent

The quaternion formulation of special relativity was introduced by 
L. Silberstein in 1912--1913 [1].  In this work, Silberstein made use of the 
original quaternion basis due to Hamilton $(1,\bfi,\bfj,\bfk)$, where $ \bfi\bfj
 = -\bfj\bfi = \bfk $ (cyclic) and ${\bfi}^2 = {\bfj}^2 = {\bfk}^2 = -1 $.
Here, we adopt the system of Pauli operators
$ (\sigma_0, \bfsigma_1, \bfsigma_2, \bfsigma_3) $, where
$ {\bfsigma}_1 {\bfsigma}_2 = -{\bfsigma}_2 {\bfsigma}_1 = i{\bfsigma}_3 $
(cyclic) and $ {\bfsigma}^2_1 = {\bfsigma}^2_2 = {\bfsigma}^2_3 = \sigma_0 $.
The fundamental interval in this theory is a quaternion, i.e., a sum of scalar
and three-vector displacements

\begin{eqnarray}
    ds & = & c\,d\tau + d{\bfr}  \nonumber \\
       & = & \sigma_0 \, dx^0 + {\bfsigma}_i \, dx^i  \nonumber \\
       & = & \sigma_\mu \, dx^\mu
\end{eqnarray}
The operator $ \sigma_0 $ commutes with all operators $ \sigma_\mu $ and,
in this way, time is distinguished as a scalar variable.  The interval $ ds $
transforms according to the general rule

\beq
   \overline{ds} =  a(ds)b
\eeq
where $a$ and $b$ are also quaternions.  For example, the Lorentz 
transformation

\begin{eqnarray}
    \overline{dx^0} & = & \gamma (dx^0 - \frac{v}{c}\, dx^1)   \\
    \overline{dx^1} & = & \gamma (dx^1 - \frac{v}{c}\, dx^0)  \\
    \overline{dx^2} & = & dx^2    \\
    \overline{dx^3} & = & dx^3
\end{eqnarray}
is given by (2), with [1,4]

\begin{eqnarray}
 a = b & = & \pm \frac{1}{\sqrt{2}}
  (\sigma_0 \sqrt{\gamma + 1} - {\bfsigma}_1 \sqrt{\gamma - 1})  \nonumber \\
  & = & \pm (\sigma_0 \cosh \frac{\xi}{2} - {\bfsigma}_1 \sinh \frac{\xi}{2})
\end{eqnarray}
and

\beq
     \gamma = \cosh \xi = \left(1 - \frac{v^2}{c^2} \right)^{-1/2}
\eeq

A scalar product can be defined by introducing the conjugate basis
${\sigtilde}_\mu = (\sigma_0, - {\bfsigma}_i) $ and conjugate interval [5]

\beq
 d\stilde = \sigtilde_\mu \, dx^\mu = \sigma_0 \, dx^0 - {\bfsigma}_i \, dx^i
\eeq
It follows from the commutation relations that

\begin{eqnarray}
    ds \, d\stilde & = & (\sigma_0  \, dx^0 + {\bfsigma}_i \, dx^i)
                     (\sigma_0 \, dx^0 - {\bfsigma}_j \, dx^j)  \nonumber \\
                   & = & \sigma_0(dx^{0^2} - dx^{1^2} - dx^{2^2} - dx^{3^2})
\end{eqnarray}
Transformers $a$ and $b$ have unit magnitude

\beq
    a\atilde = b\btilde = \sigma_0
\eeq
therefore, the scalar product is invariant

\beq
   \overline{ds\, d\stilde} = a(ds)b\,\btilde(d\stilde)\atilde = ds\, d\stilde
\eeq
It is convenient to form the symmetric products

\beq
     \frac{1}{2} (\sigma_\mu \sigtilde_\nu + \sigma_\nu \sigtilde_\mu)
       = \sigma_0 \, \eta_{\mu\nu}
\eeq
where

\beq
       \eta_{\mu\nu} = \left( \begin{array}{cccc}
                                           1&&&0 \\
                                           &-1&& \\
                                           &&-1&  \\
                                           0&&&-1
                              \end{array}
                        \right)
\eeq
Then
\begin{eqnarray}
  ds \, d\stilde & = & \sigma_\mu \sigtilde_\nu \, dx^\mu dx^\nu  \nonumber \\
        & = & \frac{1}{2}(\sigma_\mu\sigtilde_\nu + \sigma_\nu\sigtilde_\mu)
                                             \, dx^\mu dx^\nu   \nonumber  \\
            & = & \sigma_0 \, \eta_{\mu\nu}\, dx^\mu dx^\nu
\end{eqnarray}

The above formalism may be extended in a straightforward manner to variable
basis systems:
\begin{eqnarray}
     ds & = & c\,d\tau + d{\bfr}  \nonumber \\
          & = & e_0(x) \, dx^0 + {\bfe}_i(x) \, dx^i   \nonumber \\
          & = & e_\mu(x) \, dx^\mu
\end{eqnarray}
Here, we retain the scalar nature of time and the three-vector nature of space
by setting

\beq
     e_0(x) = e^0{}_0(x) \, \sigma_0
\eeq

\beq
     {\bfe}_i(x) = e^j{}_i(x) \, {\bfsigma}_j
\eeq
The conjugate interval is

\beq
          d\stilde = \etilde_\mu \, dx^\mu = e_0 \, dx^0 - {\bfe}_i \, dx^i
\eeq
Symmetric products are found to yield scalar functions

\beq
  \frac{1}{2}(e_\mu\,\etilde_\nu + e_\nu\,\etilde_\mu) = \sigma_0 \, g_{\mu\nu}
\eeq
where

\beq
         g_{\mu\nu} = \left( \begin{array}{cccc}
                                          g_{00}&0&0&0 \\
                                          0&&&  \\
                                          0&&g_{ij}&  \\
                                          0&&&
                                          \end{array}
                      \right)
\eeq
and

\begin{eqnarray}
             g_{00} & = & e^0{}_0 e^0{}_0     \\
             g_{ij}  & = & -\left[ e^1{}_i e^1{}_j + e^2{}_i e^2{}_j
                                                   + e^3{}_i e^3{}_j \right]
\end{eqnarray}
These metric
   \footnote{\small The term ``metric'' will be used, even though component
   $ g_{00} $ has no foundation in geometry.}
coefficients yield the invariant product

\begin{eqnarray}
    ds\,d\stilde & = & e_\mu \etilde_\nu \, dx^\mu dx^\nu  \nonumber  \\
                      & = & \sigma_0 \, g_{\mu\nu} \, dx^\mu dx^\nu
\end{eqnarray}

\section*{\large {III. The Structure of Space and Time}}

\indent

The basis system $ e_\mu(x) $ varies from point to point in a given manifold,
the vectors $ {\bfe}_i $ changing in both magnitude and direction.  This
behavior is expressed in terms of coefficients
\footnote{\small In the literature, Cartan's notation is often used,
$ de_\mu = e_\lambda Q^\lambda_{\mu\nu}\, dx^\nu $, even though the
$ e_\mu $ are not generally functions and the $ de_\mu $ are not differentials.
Here, we set $ de_\mu = ({\nabla}_\nu {e_\mu}) dx^\nu $ 
so that (25) follows.}
$ Q^\mu_{\nu\lambda} $

\beq
     {\nabla}_\nu e_\mu = e_\lambda Q^\lambda_{\mu\nu}
\eeq
The rate of change of the scalar basis must be of scalar form

\beq
     {\nabla}_\nu e_0 = e_0 Q^0_{0\nu} \;\; (\nu = 0,1,2,3)
\eeq

\beq
          Q^j_{0\nu} \equiv 0  \;\;  (j = 1,2,3)
\eeq
while the rate of change of the vector basis is a three-vector

\beq
     {\nabla}_\nu {\bfe}_i = {\bfe}_j Q^j_{i\nu}
\eeq

\beq
         Q^0_{i\nu} \equiv 0
\eeq
Therefore, the $ Q^\mu_{\nu\lambda} $ are generally non-symmetric in lower
indices.

Certain of the coefficients $ Q^\mu_{\nu\lambda} $ have tensor character.
These are determined in the following way.  A given observer may record a time
measurement $ d\tau $ and vector displacement $ d{\bfr} $ with respect to
basis $e_\mu $:

\beq
       c\,d\tau = e_0 \, dx^0, \;\;\; d{\bfr} = {\bfe}_i \, dx^i
\eeq
This observer is free to change the vector basis
${\bfe}_i \longrightarrow {{\bfe}_i}' $, in which case

\beq
 d{\bfr}' = {{\bfe}_i}'\,{dx^i}'\;\;\;{\mbox{\rm where}}\;\;\;d{\bfr}' = d{\bfr}
\eeq
He may also change the scalar basis $ e_0 \longrightarrow {e_0}' $ (rate of
time coordinate clocks), without affecting the length of time involved

\beq
  c\, {d\tau}' = {e_0}' \,{dx^0}' \;\;\;{\mbox{\rm where}}\;\;\;{d\tau}' = d\tau
\eeq
The new time coordinate is independent of space coordinate labels,
${x^0}' = {x^0}'(x^0)$, and the new coordinates are independent of clock rates,
$ {x^i}' = {x^i}'(x^j) $.  In short, since no relative motion is involved, the
transformation reduces to

\beq
     {e_0}' = \frac{\p x^0}{\p {x^0}'} e_0
\eeq

\beq
     {{\bfe}_i}' = \frac{\p x^j}{\p {x^i}'} {\bfe}_j
\eeq
Applying these restricted transformations to the $ Q^\mu_{\nu\lambda} $,
we find

\begin{eqnarray}
  {Q^0_{00}}' & = & \frac{\p x^0}{\p{x^0}'} Q^0_{00}
                + \frac{\p{x^0}'}{\p x^0} \frac{\p^2 x^0}{{\p x^0}^{'2}} \\
  {Q^0_{0k}}' & = & \frac{\p x^a}{\p{x^k}'} Q^0_{0a} \\
  {Q^i_{j0}}'  & = &
  \frac{\p{x^i}'}{\p x^a} \frac{\p x^b}{\p{x^j}'} \frac{\p x^0}{\p{x^0}'}
                                             Q^a_{b0} \\
  {Q^i_{jk}}' & = &
  \frac{\p{x^i}'}{\p x^a}\frac{\p x^b}{\p{x^j}'}\frac{\p x^c}{\p{x^k}'}Q^a_{bc}
              + \frac{\p{x^i}'}{\p x^a} \frac{\p^2 x^a}{\p{x^j}' \p{x^k}'}
\end{eqnarray}
Functions $Q^0_{0k} $ and $ Q^i_{j0} $ transform as tensors and thus have
invariant significance for our observer.  Defining

\beq
     Q^\mu_{[\nu\lambda]} \equiv Q^\mu_{\nu\lambda} - Q^\mu_{\lambda\nu}
\eeq
we obtain the complete array of tensor components
\begin{eqnarray}
    Q^0_{[0k]} & = & Q^0_{0k}- Q^0_{k0} = Q^0_{0k}  \\
    Q^0_{[jk]} & = & Q^0_{jk} - Q^0_{kj} = 0  \\
    Q^i_{[j0]} & = & Q^i_{j0} - Q^i_{0j} = Q^i_{j0} \\
    Q^i_{[jk]} & = & Q^i_{jk} - Q^i_{kj}
\end{eqnarray}

The vectors $ {\bfe}_i $ form the basis of a time-dependent, three-dimensional
geometry, which is characterized by metric components $g_{ij} $.  The metric
changes from point to point in the spatial manifold according to the formula

\beq
   \frac{\p g_{ij}}{\p x^k} = g_{ia} Q^a_{jk} + g_{ja}Q^a_{ik}
\eeq
By imposing nine symmetry conditions

\beq
      Q^i_{jk} = Q^i_{kj}
\eeq
this formula can be inverted in the usual way to yield

\beq
 Q^i_{jk} = \frac{1}{2} g^{ia} \left(
 \frac{\p g_{ja}}{\p x^k} + \frac{\p g_{ak}}{\p x^j} - \frac{\p g_{jk}}{\p x^a}
                                \right)
\eeq
The time-dependence of the metric is expressed in terms of coefficients
$ Q^i_{j0} $:

\beq
      \frac{\p g_{ij}}{\p x^0} = g_{ia}Q^a_{j0} + g_{ja}Q^a_{i0}
\eeq
In order to invert this formula, we impose three conditions

\beq
     g_{ia}Q^a_{j0} = g_{ja}Q^a_{i0}
\eeq
obtaining

\beq
    Q^i_{j0} = \frac{1}{2} g^{ia} \frac{\p g_{ja}}{\p x^0}
\eeq
Finally, the function $g_{00} $ changes according to

\beq
        \frac{\p g_{00}}{\p x^\mu} = 2 g_{00}Q^0_{0\mu}
\eeq
which immediately gives

\beq
     Q^0_{0\mu} = \frac{1}{2} g^{00} \frac{\p g_{00}}{\p x^\mu}
\eeq
These formulae express all 28 independent coefficients $ Q^\mu_{\nu\lambda} $
in terms of the 28 derivatives $ {\p g_{\mu\nu}}/{\p x^\lambda} $.

We can state these results more concisely by noting that conditions (48),
together with (45), are equivalent to the equation

\beq
   g_{\mu\alpha}Q^\alpha_{[\nu\lambda]} + g_{\nu\alpha}Q^\alpha_{[\lambda\mu]}
                           + g_{\lambda\alpha}Q^\alpha_{[\mu\nu]} = 0
\eeq
We now begin with the general formula

\beq
   \frac{\p g_{\mu\nu}}{\p x^\lambda}
   = g_{\mu\alpha}Q^\alpha_{\nu\lambda} + g_{\nu\alpha}Q^\alpha_{\mu\lambda}
\eeq
in the combination

\begin{eqnarray}
 & & \hspace{-.5in} \left(\frac{\p g_{\nu\mu}}{\p x^\lambda} +
 \frac{\p g_{\mu\lambda}}{\p x^\nu} - \frac{\p g_{\lambda\nu}}{\p x^\mu}\right)=
                                                                 \nonumber \\
 & & = g_{\nu\rho}Q^\rho_{\mu\lambda} + g_{\mu\rho}Q^\rho_{\nu\lambda}
  + g_{\mu\rho}Q^\rho_{\lambda\nu} + g_{\lambda\rho}Q^\rho_{\mu\nu}
  - g_{\lambda\rho}Q^\rho_{\nu\mu} - g_{\nu\rho}Q^\rho_{\lambda\mu}\nonumber \\
 & & = 2g_{\mu\rho}Q^\rho_{\nu\lambda} + g_{\mu\rho}Q^\rho_{[\lambda\nu]}
         + g_{\nu\rho}Q^\rho_{[\mu\lambda]} + g_{\lambda\rho}Q^\rho_{[\mu\nu]}
\end{eqnarray}
and then make use of (52) to find

\beq
 \left( \frac{\p g_{\nu\mu}}{\p x^\lambda} + \frac{\p g_{\mu\lambda}}{\p x^\nu}
                               - \frac{\p g_{\lambda\nu}}{\p x^\mu} \right)
  = 2g_{\mu\rho}Q^\rho_{\nu\lambda} + 2g_{\lambda\rho}Q^\rho_{[\mu\nu]}
\eeq
Raise index $ \mu $ and define the Christoffel symbols

\beq
 \Gamma^\mu_{\nu\lambda} = \frac{1}{2} g^{\mu\alpha} \left(
  \frac{\p g_{\nu\alpha}}{\p x^\lambda} + \frac{\p g_{\alpha\lambda}}{\p x^\nu}
    - \frac{\p g_{\nu\lambda}}{\p x^\alpha} \right)
\eeq
in order to obtain the formula

\beq
    \Gamma^\mu_{\nu\lambda} = Q^\mu_{\nu\lambda}
                  + g^{\mu\alpha} g_{\lambda\beta}Q^\beta_{[\alpha\nu]}
\eeq

\section*{\large {IV.  Conservation of Energy-Momentum }}

\indent

The density of energy, momentum, and stress is represented by

\beq
           T = e_\mu \otimes \etilde_\nu T^{\mu\nu}
\eeq
where $ T^{\mu\nu} = T^{\nu\mu} $.  Consider the product

\begin{eqnarray}
 (\sqrt{-g}\,T, \,dV) & = & e_\mu \sqrt{-g}\, T^{\mu\nu} \, dV_\nu \nonumber \\
 & = & e_0 \sqrt{-g}\, T^{00}\, dV_0 + e_0 \sqrt{-g}\,T^{0j}\, dV_j \nonumber\\
 & + & {\bfe}_i \sqrt{-g}\, T^{i0}\, dV_0 + {\bfe}_i \sqrt{-g}\,T^{ij}\, dV_j
\end{eqnarray}
The first term in this expression is a scalar, which gives the amount of energy
in region $ dV_0 $

\beq
     dE = e_0 \sqrt{-g}\, T^{00} \, dV_0
\eeq
while the momentum is given by the three-vector

\beq
     c \, d{\bfP} = {\bfe}_i \sqrt{-g} \, T^{i0} \, dV_0
\eeq
The remaining terms represent the flow of energy and momentum through surfaces
$ dV_j $.

The divergence theorem for energy-momentum is [6]

\begin{eqnarray}
 \oint e_\mu \sqrt{-g}\, T^{\mu\nu} \, dV_\nu
 & = & \int \left\{ e_\mu \frac{\p \sqrt{-g} \, T^{\mu\nu}}{\p x^\nu}
 + ({\nabla}_\nu e_\mu) \sqrt{-g} \, T^{\mu\nu} \right\} \,
                                                            d^4 x \nonumber \\
& = & \int e_\mu \left\{ \frac{1}{\sqrt{-g}} \frac{\p \sqrt{-g}\,
                                                         T^{\mu\nu}}{\p x^\nu}
  + Q^\mu_{\nu\lambda} \,T^{\nu\lambda} \right\} \sqrt{-g} \, d^4 x\;\;
\end{eqnarray}
Energy-momentum is conserved if

\beq
{\mbox{\rm div}}\, T^{\mu\nu} = \frac{1}{\sqrt{-g}}
                                 \frac{\p \sqrt{-g} \,T^{\mu\nu}}{\p x^\nu}
                       + Q^\mu_{\nu\lambda} \,T^{\nu\lambda} = 0
\eeq
The divergence takes on a more useful form, if we replace $ Q^\mu_{\nu\lambda} $ by
means of (57):

\begin{eqnarray}
    {\mbox{\rm div}}\, T^{\mu\nu} & = & \left\{ \frac{1}{\sqrt{-g}}
                               \frac{\p \sqrt{-g}\, T^{\mu\nu}}{\p x^\nu}
              + \Gamma^\mu_{\nu\lambda} \, T^{\nu\lambda} \right\}
              + g^{\mu\alpha}Q^\rho_{[\eta\alpha]}T^\eta_\rho \nonumber \\
    & = & T^{\mu\nu}_{;\nu} + g^{\mu\alpha} Q^\rho_{[\eta\alpha]} T^\eta_\rho
\end{eqnarray}
(A semicolon indicates the covariant derivative defined in terms of the  
Chris\-toffel symbols.)

The density of electromagnetic energy, momentum, and stress is given by

\beq
    T^{\mu\nu}_{e-m} = F^\mu{}_\alpha F^{\alpha\nu}
                     + \frac{1}{4} g^{\mu\nu} F_{\alpha\beta} F^{\alpha\beta}
\eeq
In order to investigate conservation, we calculate

\beq
{\mbox{\rm div}}\, T^{\mu\nu}_{e-m}
                = T^{\mu\nu}_{;\nu \,e-m}
                    + g^{\mu\alpha} Q^\rho_{[\eta\alpha]} T^\eta_{\rho \, e-m}
\eeq
which, by virtue of Maxwell's equation, becomes

\beq
 {\mbox{\rm div}}\, T^{\mu\nu}_{e-m}
                = F^\mu{}_\alpha J^\alpha
                    + g^{\mu\alpha} Q^\rho_{[\eta\alpha]} T^\eta_{\rho \, e-m}
\eeq
Therefore, electromagnetic energy-momentum is not conserved in regions free of
charged matter, due to the interaction term in $ Q^\rho_{[\eta\alpha]} $.
In this way, the gravitational field $ Q^\rho_{[\eta\alpha]} $ becomes
physically manifest.

Performing a similar calculation with the matter tensor

\beq
      T^{\mu\nu}_M  = \rho c^2 u^\mu u^\nu
\eeq
we find the covariant derivative

\beq
  T^{\mu\nu}_{;\nu \, M} = \rho c^2 u^\nu \frac{\p u^\mu}{\p x^\nu}
   + c^2 u^\mu \frac{1}{\sqrt{-g}} \frac{\p \sqrt{-g} \, \rho u^\nu}{\p x^\nu}
      + \rho c^2 \Gamma^\mu_{\nu\lambda} \, u^\nu u^\lambda
\eeq
The second term is zero, if rest mass is conserved.   Taken together with
(67), we obtain the divergence

\begin{eqnarray}
  {\mbox{\rm div}}\, (T^{\mu\nu}_M + T^{\mu\nu}_{e-m})
  & = &  \rho c^2 \left( u^\nu \frac{\p u^\mu}{\p x^\nu}
   + \Gamma^\mu_{\nu\lambda}\,u^\nu u^\lambda \right) + F^\mu{}_\alpha J^\alpha
                                                    \nonumber \\ 
  & + & g^{\mu\alpha} Q^\rho_{[\eta\alpha]} (T^\eta_{\rho\, M}
                                                + T^\eta_{\rho \, e-m})
\end{eqnarray}
The first expression, set equal to zero, gives the equation of motion for
charged matter

\beq
 \rho c^2 \left( u^\nu \frac{\p u^\mu}{\p x^\nu}
   + \Gamma^\mu_{\nu\lambda}\, u^\nu u^\lambda \right) + F^\mu{}_\alpha J^\alpha
         = 0
\eeq
and we are left with the gravitational interaction

\beq
  {\mbox{\rm div}}\, (T^{\mu\nu}_M + T^{\mu\nu}_{e-m})
      = g^{\mu\alpha}Q^\rho_{[\eta\alpha]}(T^\eta_{\rho\, M}
                                                + T^\eta_{\rho \, e-m})
\eeq

This interaction immediately unites electromagnetism and matter with 
gravitation.  Moreover, the interaction has tensor character, which implies that
the gravitational field must have real dynamical content (energy, momentum, 
stress) in each region of space.  This is established in the following section.

\section*{\large {V.  Gravitational Field Equations }}

\indent

In forming a Lagrangian suitable for gravitation, we begin with the invariant
expression

\beq
 \frac{c^4}{8 \pi G}\,g^{\alpha\beta}Q^\rho_{[\eta\alpha]} Q^\eta_{[\rho\beta]}
\eeq
$ (G = 6.67 \times 10^{-8}\, \mbox{\rm cm}^3/\mbox{\rm g-sec}^2) $.
The coefficient is chosen such as to yield the gravitational energy tensor

\beq
  T^{\mu\nu}_{grav} = \frac{c^4}{4 \pi G} \left\{ g^{\mu\alpha} g^{\nu\beta}
Q^\rho_{[\eta\alpha]}Q^\eta_{[\rho\beta]} -\frac{1}{2}g^{\mu\nu}g^{\alpha\beta}
Q^\rho_{[\eta\alpha]}Q^\eta_{[\rho\beta]} \right\}
\eeq
The non-zero components of $ Q^\mu_{[\nu\lambda]} $ were determined in
section III:

\begin{eqnarray}
    Q^0_{[0k]} & = & Q^0_{0k} = \frac{1}{2} g^{00} \frac{\p g_{00}}{\p x^k} \\
    Q^i_{[j0]} & = & Q^i_{j0} = \frac{1}{2} g^{il} \frac{\p g_{lj}}{\p x^0}
\end{eqnarray}
Consider the special case of a weak gravitational field,

\begin{eqnarray}
   g_{00} & = & 1 + \frac{2}{c^2} \psi  \\
   g_{ij}  & = &  -\delta_{ij}
\end{eqnarray}
where $ \psi $ is the Newtonian potential.  Substituting into (74), we obtain
the weak-field limit of $ T^{\mu\nu}_{grav} $:

\begin{eqnarray}
   T^{00}_{grav} & = & \frac{1}{8 \pi G} (\nabla \psi)^2  \\
   T^{ij}_{grav}  & = & \frac{1}{4 \pi G} \left\{ \delta^{il}\delta^{jm}
(\nabla_l \psi)(\nabla_m \psi) -\frac{1}{2}\delta^{ij} (\nabla \psi)^2 \right\}
\end{eqnarray}
$ T^{00}_{grav} $ is the positive definite energy density of the gravitational
field.  $ T^{ij}_{grav} $ is the stress tensor of Newtonian gravitation [7].
The stress is compressive along the gravitational lines of force, with tension
acting between the lines of force.

Returning to the general case, we introduce arbitrary variations of the seven
fields $ g^{\mu\nu} = (g^{00},\, g^{ij}) $ and calculate

\begin{eqnarray}
& &  \hspace{-.3in} \delta \int \frac{c^4}{8 \pi G} g^{\alpha\beta}
  Q^\rho_{[\eta\alpha]} Q^\eta_{[\rho\beta]} \sqrt{-g} \, d^4 x =  \nonumber \\
& &  = \delta \int \frac{c^4}{8 \pi G} \left\{ g^{00} Q^l_{m0}Q^m_{l0}
        + g^{lm} Q^0_{0l}Q^0_{0m} \right\}\sqrt{-g} \, d^4 x     \nonumber  \\
& &  = \delta \int \frac{c^4}{8 \pi G} \left\{   \frac{1}{4} g^{00}g^{la}g^{mb}
                      \frac{\p g_{am}}{\p x^0} \frac{\p g_{bl}}{\p x^0}
+ \frac{1}{4}g^{lm}g^{00}g^{00}\frac{\p g_{00}}{\p x^l}\frac{\p g_{00}}{\p x^m}
                                \right\}  \sqrt{-g} \, d^4 x     \nonumber \\
& &  = \int \left\{ \frac{c^4}{8 \pi G} \frac{1}{\sqrt{-g}}
   \frac{\p}{\p x^l}(\sqrt{-g}\, g^{lm} Q^0_{0m}) + \frac{1}{2} T^0_{0\, grav}
            \right\} g_{00} \delta g^{00} \sqrt{-g} \, d^4 x      \nonumber \\
& &  + \int \left\{ \frac{c^4}{8 \pi G} \frac{1}{\sqrt{-g}}
     \frac{\p}{\p x^0}(\sqrt{-g} \, g^{00}Q^i_{j0}) + \frac{1}{2} T^i_{j\,grav}
            \right\} g_{il} \delta g^{lj} \sqrt{-g} \, d^4 x
\end{eqnarray}
We have made use of these formulae

\beq
 \delta \sqrt{-g} \, = -\frac{1}{2} \sqrt{-g} \, g_{\mu\nu} \,\delta g^{\mu\nu}
\eeq

\beq
   \delta g_{\mu\nu} = -g_{\mu\alpha}\, g_{\nu\beta} \, \delta g^{\alpha\beta}
\eeq

\beq
   \delta \left( \frac{\p g_{\mu\nu}}{\p x^\lambda} \right)
            = \frac{\p}{\p x^\lambda} (\delta g_{\mu\nu})
\eeq
as well as integration by parts; variations $ \delta g^{\mu\nu} $ have been set
equal to zero at the limits of integration.  The first term in (81) will yield
Poisson's equation

\beq
    \nabla^2 \psi = 4 \pi G \, \rho
\eeq
if we introduce the matter tensor $ T^{\mu\nu}_{M} $ (68).  This is
accomplished by means of the variation

\[
  \delta\int -\left\{\rho c^2(g_{\mu\nu}u^\mu u^\nu)^{1/2}
                                    - \rho c^2\right\}\sqrt{-g} \, d^4 x
    = \frac{1}{2} \int T_{\mu\nu\,M} \delta g^{\mu\nu}\sqrt{-g} \, d^4 x
\]
\beq
  = \frac{1}{2} \int \left\{T^0_{0\,M} g_{00} \delta g^{00}
     + T^i_{j\,M} g_{il} \delta g^{lj} \right\} \sqrt{-g} \, d^4 x
\eeq
The final expression to be introduced is the energy tensor of electromagnetism
[8]

\[
  \delta \int -\frac{1}{4} F_{\mu\nu}F^{\mu\nu} \sqrt{-g} \, d^4 x
   = \frac{1}{2} \int T_{\mu\nu\,e-m} \delta g^{\mu\nu} \sqrt{-g}\, d^4 x
\]

\beq
  = \frac{1}{2} \int \left\{ T^0_{0\,e-m} g_{00} \delta g^{00}
   + T^i_{j\,e-m} g_{il} \delta g^{lj} \right\} \sqrt{-g} \, d^4 x
\eeq
where $ T^{\mu\nu}_{e-m} $ is given by (65).

Gathering terms together and setting coefficients of $ \delta g^{00} $ and
$ \delta g^{ij} $ equal to zero, we arrive at the field equations of 
gravitation

\beq
 \frac{c^4}{4 \pi G} \frac{1}{\sqrt{-g}}
                     \frac{\p}{\p x^l}(\sqrt{-g}\, g^{lm} Q^0_{0m})
           +  T^0_0 = 0
\eeq

\beq
  \frac{c^4}{4 \pi G} \frac{1}{\sqrt{-g}}
                      \frac{\p}{\p x^0}(\sqrt{-g} \, g^{00} Q^i_{j0})
             + T^i_j = 0
\eeq
$ T^{\mu\nu} $ represents the total energy, momentum, and stress

\beq
  T^{\mu\nu} = T^{\mu\nu}_{grav} + T^{\mu\nu}_M + T^{\mu\nu}_{e-m}
\eeq
The corresponding invariant equation is

\beq
  \frac{c^4}{4 \pi G} \frac{1}{\sqrt{-g}}
                      \frac{\p}{\p x^\nu}(\sqrt{-g}\, Q^\nu)
               + T = 0
\eeq
where

\beq
          Q_\mu \equiv Q^\alpha_{[\alpha\mu]}
\eeq

\section*{\large {VI. The Gravitational Field of an Electron}}

\indent

According to our hypothesis, it is the electromagnetic nature of matter which
gives rise to gravitation.  Thus, our central problem is to determine the field
of a discrete charged particle, at rest.  The field equations admit an exact
solution for this case, of the form

\beq
          g_{\mu\nu} = \left( \begin{array}{cccc}
                                          e^\nu & & & 0 \\
                                          &-e^\lambda & &  \\
                                          & & -r^2 &  \\
                                          & 0 & & -r^2 \sin^2 \theta
                                           \end{array}
                                  \right)
\eeq
where $ \nu $ and $ \lambda $ are functions of $ r $ which vanish at infinity.
The electromagnetic energy tensor is [9]

\beq
    T_{\mu\nu \, e-m} = \frac{q^2}{8 \pi r^4}\left( \begin{array}{cccc}
                                                    e^\nu & & & 0 \\
                                                    & -e^\lambda & & \\
                                                    & & r^2 &  \\
                                                    & 0 & & r^2 \sin^2 \theta
                                                    \end{array}
                                             \right)
\eeq
where $ q $ is the electric charge in ordinary c.g.s. units (stat-coulomb).
The components of the gravitational energy tensor are found from (74) and
(93) to be

\beq
    T_{\mu\nu \, grav} = \frac{c^4}{32 \pi G} e^{-\lambda} {\nu'}^2
                                            \left( \begin{array}{cccc}
                                                   e^\nu & & & 0 \\
                                                   & e^\lambda & & \\
                                                   & & -r^2 & \\
                                                   & 0 & & -r^2 \sin^2 \theta
                                                   \end{array}
                                            \right)
\eeq
(A prime indicates differentiation with respect to $ r $).  Finally, we have

\begin{eqnarray}
    & & \hspace{-.8in}\frac{\p }{\p x^l} (\sqrt{-g} \, g^{lm} Q^0_{0m}) 
            = \frac{d}{dr} (\sqrt{-g} \, g^{11} Q^0_{01}) = \nonumber \\
    & & = - \frac{1}{2} r^2 \sin\theta e^{(\nu - \lambda)/2} \left(
    \nu'' + \frac{2}{r} \nu' + \frac{1}{2} {\nu'}^2 - \frac{1}{2} \nu'\lambda'
                 \right)
\end{eqnarray}

Substitution into (88) and (89) yields two distinct equations

\beq
    \nu'' + \frac{2}{r} \nu' + \frac{1}{4}{\nu'}^2
    - \frac{1}{2} \nu' \lambda' - \frac{Gq^2}{c^4 r^4} e^\lambda = 0
\eeq

\beq
   - \frac{1}{4} {\nu'}^2 + \frac{Gq^2}{c^4 r^4} e^\lambda = 0
\eeq
Adding the two equations gives

\[
     \nu'' + \frac{2}{r} \nu' - \frac{1}{2} \nu' \lambda' = 0
\]
or

\beq
    d\,(\ln \nu') + \frac{2}{r}\, dr = \frac{1}{2}\, d\lambda
\eeq
Upon integration, we find an equation of the form (98).  In order to arrive at
a unique solution, we set
\beq
   \lambda = - \nu
\eeq
That is, we invoke a reciprocal relationship between time dilatation and length
contraction along the direction of the field.  Equation (98) now gives

\beq
   e^{\nu /2} \, d\nu = \frac{2G^{1/2} q}{c^2 r^2} \, dr
\eeq
This equation integrates to

\beq
  e^\nu = e^{-\lambda} = \left( 1 - \frac{G^{1/2} q}{c^2 r} \right)^2
\eeq
the constant of integration being chosen such that $ e^\nu = e^{-\lambda} = 1 $
at infinity.  The components of the metric tensor are

\beq
   g_{\mu\nu} = \left( \begin{array}{cccc}
            \left( 1 - \disp\frac{G^{1/2} q}{c^2 r} \right)^2 & & & 0 \\
            & - \left( 1 - \disp\frac{G^{1/2} q}{c^2 r} \right)^{-2} & & \\
            & &  -r^2 &  \\
            0 & & & -r^2 \sin^2 \theta
                       \end{array}
                \right)
\eeq

The gravitational field of an electron is given by setting
$ q = e = 4.8 \times 10^{-10} {\mbox{\rm stat-coulomb}} $ in (103). Noting that
\beq
     G^{1/2} e/c^2  = 1.4 \times 10^{-34} \, {\mbox{\rm cm}}
\eeq
the field may be represented by  a gravitational potential

\beq
       \psi = -G^{1/2} \frac{e}{r}
\eeq
This potential is some 21 orders of magnitude greater than the Newtonian value,
$ -Gm/r $.  It acts in those regions of space which are subject to the 
electrostatic field of the electron.

\section*{\large {VII.  An Electrostatic Red-Shift in Stars}}

\indent

In the previous section, it was shown that a spherically symmetric     
electrostatic field produces a gravitational potential given by

\beq
        \psi = -G^{1/2} \frac{q}{r}
\eeq
where $ q $ is the absolute value of electric charge.  This potential, in turn,
will produce a red-shift of light given by

\beq
   \frac{\Delta \nu}{\nu_0} = - \frac{1}{c^2} \psi = \frac{G^{1/2} q}{c^2 r}
\eeq
where $ \nu_0 $ is the light frequency in the absence of gravitation, i.e., at
$ r \longrightarrow \infty $.  In this section, we discuss a plausible 
observation of this red-shift and its correlation with ordinary electrodynamic
processes.

For purposes of our discussion, consider the following hypothetical stellar
object, in which a spherically symmetric charge separation has occurred.
Neutrons and protons, held strongly in the core, are surrounded by electrons
which have been forced away by thermal energy and mutual repulsion.  Under the
simplest assumption, all electrons lie in a spherical shell of fixed radius.
This model may be treated in terms of our theory as follows.  Each neutron
produces a Newtonian potential

\beq
   \psi_n = - \frac{Gm}{r} \doteq -\frac{10^{-31}}{r}
\eeq
$(m = 1.7\times 10^{-24} {\mbox{\rm g}})$, while each proton, according to (106)
produces the much greater potential

\beq
   \psi_p = -G^{1/2} \frac{e}{r} \doteq -\frac{10^{-13}}{r}
\eeq
Ignoring the neutron contribution, we write the gravitational potential (106)
in the form

\beq
   \psi = - G^{1/2} \phi
\eeq
where $ \phi $ is the total electrostatic potential.  A ray of light, being
emitted from the core and passing through potential $ \phi $, will suffer a
red-shift (107) given by

\beq
      \frac{\Delta \nu}{\nu_0} = \frac{G^{1/2}}{c^2} \phi
\eeq
This linear relation holds true over a very wide range of electrostatic
potential (see below).  In addition to light, one would expect positively
charged particles (cosmic rays) to be emitted from such a stellar object.
Particle energy is commensurate with the electrostatic potential.  Thus,
according to (111), one should observe a linear correlation between the
red-shift of light and the energy of cosmic rays.

The exact solution to the electrostatic problem (103) takes the form

\beq
 g_{\mu\nu} = \left( \begin{array}{cccc}
              \left( 1 - \disp\frac{G^{1/2}}{c^2} \phi \right)^2 & & & 0 \\
              & - \left( 1 - \disp\frac{G^{1/2}}{c^2} \phi \right)^{-2} & & \\
              & & -r^2 & \\
              0  & & & -r^2 \sin^2 \theta
                     \end{array}
              \right)
\eeq
Component $ g_{00} $ gives the frequency relation

\beq
 \frac{\nu}{\nu_0} = g^{-1/2}_{00}
                   = \left( 1 - \frac{G^{1/2}}{c^2} \phi \right)^{-1}
\eeq
and red-shift formula

\beq
  \frac{\Delta \nu}{\nu_0} = \frac{\disp\frac{G^{1/2}}{c^2} \phi}
                                      {1 - \disp\frac{G^{1/2}}{c^2} \phi }
\eeq
According to this formula, an infinite red-shift corresponds to a potential of

\begin{eqnarray}
   c^2/G^{1/2} & = & 3.5 \times 10^{24} \,{\mbox{\rm stat-volts}} \nonumber \\
                        & = & 10^{27} \, {\mbox{\rm volts}}
\end{eqnarray}
It follows that cosmic ray energy can be no greater than $ 10^{27} $ eV.

It would be of interest to search for evidence of a correlation between the
gravitational red-shift of a discrete source and the energy of cosmic rays
emitted by that source.  This may be within reach of modern experimental
methods in astronomy.  Perhaps of greater significance is the predicted upper
limit to cosmic ray energy, $ 10^{27} $ eV.  This limit is associated with an
infinitely red-shifted stellar object, an ``electrostatic black-hole,'' at a
potential of $ 10^{27} $ volts.  The astronomers have found the maximum energy
of cosmic rays to be at least $ 10^{20} $ eV.  This compares favorably with the
theoretical value.

\section*{\large {VIII. Concluding Remarks }}

\indent

According to equation (89), any flow of momentum (i.e., stress) will give
rise to a time-dependent gravitational field.  This applies, in particular,
to the field generated by a stationary uncharged mass such as the Sun.
In a first order approximation to this problem, all stress terms may be
ignored; (88) then yields Poisson's equation and the static Newtonian
potential.  However, in second order, the theory predicts the existence of 
time-dependent perturbations in the Newtonian field, due to the presence of 
material and gravitational stress.  Recently discovered resonant oscillations 
in the Sun show that the solar field is, indeed, time-dependent.  It remains to
be seen whether this provides corroboration for our theory.

The above behavior contrasts with that of the field surrounding a stationary
electric charge (section VI).  Here, the electrostatic stresses are equal and
opposite to those of the gravitational field.  Therefore, the net flow of
momentum is everywhere zero, and the solution is time-independent.  This result
underscores the significance of electric charge in our theory.  Moreover,
it calls to mind the central role played by electric charge and `generalized
charge' in elementary particle theory.  The enormous strength of the electron's
gravitational potential may point toward an eventual link with the physics of
elementary particles.

\vspace{.5in}

\section*{\large {References}}

\begin{enumerate}

\item L. Silberstein, {\it Phil.Mag.} {\bf 23}, 790 (1912); {\bf 25}, 135 (1913).
\item The use of these operators in three-dimensional rotations is discussed by
   H.~Corben and P.~Stehle, {\it Classical Mechanics} (Krieger,Huntington, NY,
   2nd ed. 1974) p.~373--381.  Also, D.~Hestenes, {\it Am. J. Phys.} {\bf 39},
   1013 (1971).
\item The equivalent spin-matrix approach to rotations is given by H.~Goldstein,
{\it Classical Mechanics} (Addison-Wesley, 1950) p.~109--118;
C.~Misner, K.~Thorne, J.~Wheeler, {\it Gravitation} (Freeman, New York, 1973)
p.~1135; E.~Sudarshan and N.~Mukunda, {\it Classical Dynamics: A Modern
Perspective} (Krieger, Malabar FL, 1983) p.~345--348; E.~Gottfried and
V.~Weisskopf, {\it Concepts of Particle Physics} (Oxford, 1986) vol.~II, p.~559.
\item The Lorentz transformation is given in spin-matrix form by C.~Misner, 
    K.~Thorne, J.~Wheeler (Ref. 3), p.~1142; E.~Sudarshan, N.~Mukunda (Ref. 3), 
    p.~480--484; K.~Gottfried, V.~Weisskopf (Ref. 3), p.~560--561.
\item A. Barut, {\it Electrodynamics and Classical Theory of Fields and 
      Particles} (Macmillan, New York, 1964), pp.~25, 43.
\item K. Dalton, {\it Gen.Rel.Grav.} {\bf 21}, 533 (1989).
\item C. Misner, K. Thorne, J. Wheeler (Ref. 3), pp. 470, 1078.
\item S. Weinberg, {\it Gravitation and Cosmology} (Wiley, New York, 1972)
       p.~360; L.~Landau and E.~Lifshitz, {\it The Classical Theory of Fields} 
       (Pergamon, 4th ed., 1975) p.~272.
\item R. Adler, M. Bazin, M. Schiffer, {\it Introduction to General Relativity}
       (McGraw-Hill, New York, 1965) sect.~13.1.

\end{enumerate}

\end{document}